\documentclass[12pt,preprint]{aastex}

\def\psr{PSR J0108--1431\/}

\shorttitle{Optical counterpart of PSR J0108--1431} 
\shortauthors{Mignani, Manchester, \& Pavlov}
\begin{document}

\title{Search for the optical counterpart of the nearby pulsar 
J0108-1431.\footnote{Based on observations 
obtained with ESO under Progr.\ 65.H-0400(A)}
}

\author{Roberto P.\ Mignani\footnote{
European Southern Observatory, Karl-Schwarzschild-Str.\ 2, 
D85740, Garching, Germany;
rmignani@eso.org},
Richard N.\ Manchester\footnote{
Australia Telescope National Facility, CSIRO, P.O.\ Box 76, 
Epping, NSW 1710, Australia;
dick.manchester@csiro.au}, and
George G.\ Pavlov\footnote{
Pennsylvania State University, 525 Davey Lab, University Park, PA 16802, USA;
pavlov@astro.psu.edu}}

\begin{abstract}
We present the results of first deep optical observations of the field
of the old ($\sim 10^8$ yr), nearby, isolated pulsar J0108$-$1431, in
an attempt  to detect its optical counterpart.   The observations were
performed  using the  FORS1 instrument  at the  focus of  the European
Southern Observatory Antu Telescope  of the VLT. Observations with the
Australia Telescope  Compact Array (ATCA)
 were made to  determine an accurate
position for the radio pulsar  at the current epoch. The imaging data,
obtained in the  $V$, $B$, and $U$ passbands  reveal no counterpart at
the revised radio  position down to $V \simeq  28$, $B\simeq 28.6$ and
$U  \simeq  26.4$.  For  a  distance of  130  pc,  estimated from  the
pulsar's dispersion  measure, our constraints on the  optical flux put
an upper limit of $T=4.5\times  10^4$ K for the surface temperature of
the neutron star, assuming a stellar radius $R_\infty=13$ km.  Our new
radio position allows us to place  an upper limit on the pulsar proper
motion of 82~mas~yr$^{-1}$ which, for $d=130$ pc, implies a
transverse velocity $\la 50$~km~sec$^{-1}$.
\end{abstract}


\section{Introduction}

PSR  J0108$-$1431  was discovered  by  Tauris  et  al.\ (1994)  during
observations performed as a part  of the Parkes Southern Pulsar Survey
(Manchester et  al.\ 1996).   With a period  $P=0.808$ s and  a period
derivative  $\dot{P} =  7.44 \times  10^{-17}$ s~s$^{-1}$  (D'Amico et
al.\ 1998),  the pulsar  has a characteristic  age $P/2\dot{P}  = 170$
Myr,  rotation  energy   loss  rate  $\dot{E}=5.6\times  10^{30}$  erg
s$^{-1}$, and magnetic field  $B=2.5\times 10^{11}$ G.  The dispersion
measure of  $2.38 \pm 0.01$  cm$^{-3}$~pc (D'Amico et al.\  1998), the
lowest observed so far for a radio pulsar, puts \psr\ at a distance of
about 130 pc, estimated according to the Taylor \& Cordes (1993) model
for the Galactic electron  density distribution.  Tauris et al.\ (1994)
suggest that the local electron density may be greater than average in
this direction, giving a distance as  low as 60 pc.  Regardless of the
exact value,  the small distance  to PSR J0108$-$1431 implies  a radio
luminosity  a factor  $10^3$ smaller  than the  average for  other old
pulsars of similar ages.  This  pulsar could thus be representative of
a population of sub-luminous radio pulsars (e.g., Dewey et al. 1985).

Since \psr\  is apparently the closest  known pulsar, it  is a natural
candidate for observations in other wavelength bands.  However, it has
not yet been  detected outside the radio range.   Presumably too faint
to be detected  in the {\sl ROSAT} All Sky Survey,  \psr\ has not been
observed in  a pointing  mode with an  X-ray observatory so  far.  The
field was  observed with the {\sl Extreme  Ultraviolet Explorer} ({\sl
EUVE}), but the pulsar was  not detected (Korpela \& Bowyer 1998).  In
the   mid-infrared,  observations   with  the   {\sl   Infrared  Space
Observatory} were performed to search for a disk around the pulsar but
with no conclusive result  (Koch-Miramond et al.\ 2002).  Particularly
interesting  would be  the detection  of optical-UV  thermal radiation
from  the  neutron  star  (NS)  surface  to  examine  various  heating
mechanisms  that can  operate in  the NS  interiors.  According  to the
current models (see, e.g., Tsuruta 1998,  for a review), by the age of
170  Myr  a   NS  would  have  cooled  down   to  very  low  (surface)
temperatures,  $T  < 10^4$  K,  and  its  thermal radiation  would  be
virtually undetectable unless some (re)heating mechanisms operate.

Amongst several proposed heating mechanisms, two are most efficient in
slowly rotating pulsars.   The first one is the  dissipation of energy
of differential  rotation.  As invoked  by models of  pulsar glitches,
the interior  neutron superfluid rotates  more rapidly than  the outer
solid  crust.  Frictional  interaction with  the crust  dissipates the
energy of  the differential rotation,  heating the star  (Shibazaki \&
Lamb 1989;  Umeda et al.~1993; Van  Riper et al.~1994;  Larson \& Link
1999).   The  amount of  heat  released  depends  on the  differential
angular momentum  $\Delta J_s$ of the  frictionally coupled superfluid
layers.  For instance, at an  age $\tau\sim 200$ Myr, plausible values
of  $\Delta  J_s$  can   provide  $T\simeq  (3$--$10)\times  10^4$  K,
depending on the equation of  state of the NS interiors and properties
of  nucleon  superfluidity.   The  other heating  mechanism  is  Joule
heating caused by  dissipation of the magnetic field  in the NS crust.
According to Miralles, Urpin \&  Konenkov (1998), the NS luminosity at
$\tau \gtrsim$  10 Myr is  approximately equal to the  energy released
due to the field dissipation.   At this stage, the surface temperature
decreases slowly, being comparable  to that produced by the frictional
heating at $\tau \sim 100$ Myr.

Thus, if  the heating  mechanisms indeed operate,  one can  expect the
surface  NS  temperature of  a  few  times  $10^4$~K, which  would  be
detectable  in  the  optical-UV  range  but  undetectable  in  X-rays,
traditionally used for observations of thermal radiation from NSs.  It
is  also possible
that  the  pulsar emits  nonthermal
optical radiation.   So far, this  has only been firmly  detected from
younger pulsars  (e.g., the Vela  pulsar --- Mignani \&  Caraveo 2001;
PSR B0656+14 --- Koptsevich et al.\ 2001).

The  only  previous  observations   of  \psr\  in  the  optical  range
($B$,$V$,$R$,$I$  bands) were carried  out with  the 6-m  telescope of
Special  Astrophysical Observatory  (Russia), but  they were  not deep
enough  to detect  the pulsar  and provide  useful constraints  on the
radiation mechanisms  (Kurt et al.\ 2000).  Therefore,  we performed a
deep observation  of the pulsar field  with one of the  ESO Very Large
Telescopes (VLT) units in several passbands.

The most recent published  coordinates of PSR J0108$-$1431 (D'Amico et
al.\ 1998) were derived from timing data obtained in the interval 1993
April  to 1996 June,  giving a  mean epoch  of 1994.9.  Therefore, our
knowledge of the actual position of the pulsar at the epoch of our VLT
observations  is affected  by  a significant  uncertainty  due to  its
unknown proper motion.   A pulsar distance of 130  pc and a transverse
velocity  of 400  km  s$^{-1}$, a  typical  value for  a radio  pulsar
(Lorimer,  Bailes \&  Harrison 1997),  give a  proper motion  of $\sim
0\farcs65$ yr$^{-1}$.  For the epoch of our VLT observations (2000.6),
such  a  proper  motion  would  imply a  displacement  in  an  unknown
direction of $\sim 3\farcs5$ with respect to the original radio timing
position.    This  obviously   makes   impossible  a   straightforward
positional search  for the optical  counterpart.  For this  reason, we
observed the pulsar with  the Australia Telescope Compact Array (ATCA)
to obtain  a radio position at an  epoch close to that  of our optical
observations and to measure the pulsar's proper motion.

We  describe the  radio and  optical  observations in  \S2, while  the
results and their implications are  presented and discussed in \S3 and
\S4, respectively.

\section{Observations}

\subsection{Radio imaging}

The  pulsar  field was  observed  using  the  ATCA, a  6-km  east-west
synthesis array consisting of six 22m-diameter antennas, on 2001 March
31 for  11.3 hours. Data  were obtained in  two bands, each  with dual
polarization and bandwidth  of 128 MHz, centered on  1384 MHz and 2496
MHz, respectively. The flux density scale was calibrated using a short
observation of  the standard  source 1934-638. Two  phase calibrators,
0048-097 and  0114-211, were each observed  for 2 min every  30 min to
establish the position reference  frame. Each of these calibrators has
positional uncertainties of $\la  0\farcs1$ in the NRAO VLA calibrator
list. Pulsar-gating  mode was  used, giving 32  phase bins  across the
pulsar  period at  both  frequencies. The  128-MHz  bandwidth at  each
frequency  was  also  split  into  32  channels  allowing  removal  of
dispersive delays during processing.

Data  were  processed  using  standard techniques  with  the  analysis
package
MIRIAD\footnote{http://www.atnf.csiro.au/computing/software/miriad/}.
Use of  pulsar binning  allowed images to  be made using  on-pulse
minus
off-pulse  difference   
visibility data.  This   removed  essentially  all
artifacts from the images, allowing clear identification of the pulsar
in both  bands. Positions were derived  by fitting to the  peak of the
source, at  1384 MHz and,  separately for calibration with  each phase
calibrator, at 2496 MHz.

\subsection{Optical imaging}

Optical observations  of the field of PSR  J0108$-$1431 were performed
in  Service Mode  in  July and  August  2000 with  the 8.2-meter  Antu
Telescope of the ESO VLT,  located at the Paranal Observatory (Chile).
The images  were obtained using  the FOcal Reducer and  Spectrograph 1
(FORS1) instrument,  a four-port $2048\times 2084$  CCD detector which
can be used both as  a high/low resolution spectrograph and an imaging
camera.  The instrument  was operated in imaging mode  at its standard
angular resolution  of 0\farcs2 per pixel, with  a corresponding field
of  view of  $6\farcm8 \times  6\farcm8$.  We  chose to  sacrifice the
angular resolution in favor of a  larger field of view to increase the
number of reference stars needed  for a precise CCD astrometry (see \S
2.3).  The  images were taken in  the Bessel filters      
$U$    ($\lambda=3660$\AA,
$\Delta\lambda=360$\AA),             $B$             ($\lambda=4290$\AA,
$\Delta\lambda=880$\AA),        and        $V$       ($\lambda=5540$\AA,
$\Delta\lambda=1150$\AA).   For   each  of  the   filters,  the  total
integration time  was split in  shorter exposures to  avoid saturation
effects of  the CCD and allow  for a better cosmic  ray filtering.  In
five nights we collected 12 exposures  of 600~s each in $V$, 6 exposures
of 900~s in $B$, and 5 exposures of 1\,800~s in $U$, for total integration
times of  7\,200, 5\,400 and  9\,000 s, respectively.  The  journal of
the  observations is  summarized in  Table 1.   The target  was always
observed  close to  the  minimum airmass  ($\langle\sec z\rangle  \sim
1.13$) and under sub-arcsecond  seeing conditions ($\sim 0\farcs6$, as
measured directly on the image).  Atmospheric conditions were optimal,
with little wind and humidity  always below 10\%.  All the nights were
photometric,  with  the only  exception  of  the  last one  which  was
affected by the presence of variable thin cirrus.  However, the effect
of the extinction was estimated to be below 0.05 magnitudes.

Basic reduction steps to remove instrumental signatures
and to renormalize the counts for the different gains of the four CCD
ports have been applied through the FORS1 reduction pipeline.  For each
night, master bias and flats collected in day/night time have been
used for debiassing and flatfielding.  Finally, all the exposures
corresponding to a given observation block (see Table 1) have been
combined using a median filter algorithm to remove cosmic ray hits.
The $U$ and $V$ band images taken in different nights were further
registered on each other and coadded.  For all the science images, the
flux calibration was performed using images of photometric standards
obtained during the same nights, yielding extinction and
color-corrected zero-points with an accuracy of a few hundredths of
magnitude for each of the filters. The atmospheric extinction
correction was computed according to the most recent coefficients
measured for Paranal\footnote{http://www.eso.org/observing/dfo/quality/FORS1/qc/photcoeff/}. The zero-points were stable,
showing negligible fluctuations from night to night, which makes it
possible to compare the photometry of different nights.

\section{Results}

\subsection{Pulsar position and proper motion}

The ATCA observations give a  mean position in J2000 coordinates, with
estimated  uncertainties, of  R.A.=$01^{\rm  h}08^{\rm m}08\fs  317\pm
0\fs010$, Dec.=$-14\degr  31\arcmin 49\farcs35\pm 0\farcs35$  at epoch
2001.3   (MJD  51999).   This  position   is  coincident   within  the
uncertainties with that obtained from the timing data (D'Amico et al.\
1998): R.A.=$01^{\rm h}08^{\rm m}08\fs 318\pm 0\fs 005$, Dec=$-14\degr
31\arcmin 49\farcs  2\pm 0\farcs1$, which  has mean epoch  1994.9 (MJD
$\sim 49657$). These results imply an upper limit on the pulsar proper
motion  of $\mu_\alpha  < 26$  mas  yr$^{-1}$, $\mu_\delta  < 78$  mas
yr$^{-1}$.

The  corresponding  upper limits  on  the  pulsar transverse  velocity
components are $v_\alpha < 16\, d_{130}$ km s$^{-1}$, $v_\delta < 48\,
d_{130}$ km s$^{-1}$, where $d_{130}$  is the pulsar distance in units
of 130 pc.  The total transverse velocity amounts to  $v < 50 d_{130}$
km   s$^{-1}$.  This   is   within  the   bottom   10\%  of   measured
velocities. The pulsar is at a Galactic latitude of $-77\degr$, and it
is quite possible  that its main velocity component  is along the line
of sight.  Indeed, there is some  indication of a  secular increase in
the  dispersion   measure,  suggesting  a   significant  line-of-sight
velocity component.

\subsection{The search for the optical  counterpart}

In order  to register  accurately the most recent pulsar  position on  the optical
images,  we  needed an  astrometric  solution  more  precise than  the
default one derived from the  telescope pointing, which can show frame
to frame variations up to few  arcseconds.  Thus, astrometry on the FORS1
images  was recomputed  using as  a reference  the positions  of stars
selected from the recently  released Guide Star Catalogue II 
(GSC-II\footnote{http://www-gsss.stsci.edu/gsc/gsc2/GSC2home.htm}),
which  have  an  intrinsic   absolute  astrometric  accuracy  of  $\le
0\farcs5$ per  coordinate.
About 50  GSC-II objects
have been  identified in one  of the two  averaged $V$-band  images.  Of
these, only 21  (those which are not extended, not  too faint, and not
too close to the CCD  edges) have been considered suitable astrometric
calibrators.  The  pixel coordinates of the reference  stars have been
computed by a gaussian fitting  procedure using the tools available in
the SKYCAT image display  interface.  The transformation from pixel to
sky coordinates was then computed using the program 
ASTROM\footnote{http://star-www.rl.ac.uk/Software/software.htm},
yielding  an  rms  error  of  $\sim 0\farcs2$  in  both
coordinates.

Considering the  uncertainties in both  the radio position and  in the
optical astrometry,  the final rms uncertainty in  registration of the
pulsar on the optical image is $0\farcs33$ in R.A.  and $0\farcs46$ in
Dec.   The final error  budget takes  into account  the errors  of the
radio  pulsar coordinates  at the  epoch of  the  optical observations
($0\farcs16$  and  $0\farcs36$  in  the R.A.\  and  Dec.\ directions,
respectively), the rms error  of our astrometric fit ($0\farcs2$), and
the  propagation  of  the  intrinsic  absolute errors  on  the  GSC-II
coordinates ($0\farcs2$).   We note that  since the astrometry  of the
GSC-II was  originally calibrated using  stars from the  Hipparcos and
Tycho  catalogues, which are  tied to  the extragalactic  radio source
frame (ICRF),  our position should  not be affected by  any systematic
offset between the radio and optical reference frames.

The pulsar position at the epoch of our optical observations is marked
in Figures  1 through  3, where we  show $25\farcs6  \times 25\farcs6$
cutouts  of  the co-added  $V$,  $B$,  and $U$  images.   We  note that  our
determination of the pulsar position  differs by about $5''$ from that
shown in  Figure 5  of Kurt et  al.  (2000).   The reason for  this is
probably due to  the fact that Kurt et al. assumed  as a reference the
original pulsar coordinates reported by Tauris et al. (1994).

To facilitate the object detection,  all the images have been smoothed
using a  gaussian filter with a $\sigma$  of 3 pixels in  both $X$ and
$Y$ directions.   We see  from these images  that the  computed pulsar
position  falls  close to  an  elliptical  galaxy  in the  field.  The
distance  from the  galaxy  edge is  about  $0\farcs 6$,  in the  East
direction.   Since such  a  distance corresponds  to  about twice  the
uncertainty of  the pulsar  position in R.A,  it is unlikely  that the
pulsar is hidden behind the galaxy,  but we cannot rule it out.  While
no point source is obviously visible at the pulsar location in the $V$
and  $B$ images,  an excess  of counts  can be  recognized in  the $U$
image,  with  a  magnitude   of  $\approx  27.2$.   However,  the  low
significance of  this detection ($\sim  2 \sigma$), together  with the
lack  of  corresponding  detections  in  the  other  filters  and  the
proximity of  the galaxy, suggest that  it may be due  to a background
fluctuation.  Only  few point-like objects (labelled  in Figures 1--3)
have  been detected  within a  radius of  $6''$  from  the pulsar
position.  These  objects, together  with their positions  relative to
the  pulsar  and  the  extinction-corrected photometry  in  the  three
filters, are  listed in Table 2.  However, as it is  evident from both
Table 2 and  Figures 1--3, all these objects are  too distant from the
nominal pulsar position to be considered candidate counterparts.

Therefore, we conclude  that no optical counterpart to  the pulsar can
be identified  in our data.   Using as a  reference the fluxes  of the
faintest objects detected in the pulsar surroundings (Table 2), we can
derive $3 \sigma$ upper limits of $V \simeq 28$, $B\simeq 28.6$ and $U
\simeq 26.4$ on the pulsar magnitudes. Thus our limits are deeper than
those obtained by Kurt et al.\ (2000) by more than 3 magnitudes.

\section{Discussion}


The upper limits on spectral flux at effective frequences of the three
bands (Fukugita, Shimasaku, \& Ichikawa 1995) are plotted in Figure 4.
The deepest $B$-band limit corresponds to the  following limit on the
(brightness) temperature

\begin{equation}
T_\infty<\frac{3.30\times   10^4~{\rm   K}}{\ln\left(1+
1.08\, R_{13}^2
d_{130}^{-2}\right)}\,,
\end{equation}
where $R_{13}=R_\infty/(13~{\rm km})$, $d_{130}=d/(130~{\rm pc})$, the
subscript $\infty$ denotes the quantities as measured by a distant
observer: $T_\infty=Tg_r$, $R_\infty=R/g_r$, $g_r=(1-2GM/Rc^2)^{1/2}$
is the gravitational redshift parameter ($g_r=0.769$ for a neutron
star mass $M=1.4 M_\odot$ and radius $R=10$ km.)  In this estimate, we
neglect corrections due to interstellar extinction as they are
expected to be negligible at the small distance.  For the dispersion
distance $d=130$ pc and radius $R_\infty = 13$ km, we obtain $T_\infty
< 4.5\times 10^4$ K. This value is considerably lower than $T_\infty  
\la 1.3\times 10^5$ K, which follows from the upper limit on
the {\sl EUVE} flux of this pulsar (Korpela \& Bowyer 1998), for the
same $d$ and $R_\infty$, and $N_{\rm H} < 3\times 
10^{19}$ cm$^{-2}$.  It is also much lower than the lowest upper
limit, $T_\infty\la 3\times 10^5$ K,
estimated from optical/UV observations of 
another nearby
pulsar, B0950+08~\footnote{The
limit of $7\times 10^4$ K in the paper of Pavlov et al.\ (1996) was
estimated for $d=127$ pc.  The distance $d=262\pm 5$ pc was obtained
by Brisken et al.\ (2002) from improved parallax measurements.}    
--- see Pavlov,
Stringfellow, \& C\'ordova (1996).  The limiting temperature is even
lower for $d=60$ pc, suggested by Tauris et al.\ (1994): $T<1.8\times
10^4$ K. We plot the blackbody spectra at these values of $T_\infty$
and $d$ in Figure 4.  The corresponding limits on bolometric
luminosity, $L_{{\rm bol},\infty}=L_{\rm bol} g_r^2 =4\pi R_\infty^2
\sigma T_\infty^4$, are 
$4.9\times 10^{27}$ and $1.4\times 10^{26}$
erg s$^{-1}$, for $d=130$ pc and 60 pc, respectively.
We adopt the more conservative 130 pc estimate in the following
discussion and consider $T_\infty < 4.5\times 10^4$ K and $L_{{\rm bol},
\infty} < 5\times 10^{27}$ erg s$^{-1}$ as plausible upper limits.

The limit on thermal emission strongly constrains possible heating
mechanisms. For the frictional heating mechanism (see \S1), one can
constrain the excess angular momentum $\Delta J_s$, residing in the
superfluid, and the angular velocity lag $\bar{\omega}$, between the
superfluid and the crust, averaged over the superfluid moment of
inertia. For instance, using equation (9) of Larson \& Link (1999), we
obtain $\Delta J_s=L_{{\rm bol},\infty}/|\dot{\Omega}| < 
7\times 10^{42}$ erg s, $\bar{\omega} = \Delta J_s/I_s < 
1$ rad s$^{-1}$,
where $\dot{\Omega} = -7.16\times 10^{-16}$ rad s$^{-2}$ is the time
derivative of the angular frequency of the pulsar, $I_s=7.3\times
10^{43}$ g cm$^2$ is the moment of inertia of the portion of the
superfluid that is differentially rotating, estimated for the Friedman
\& Pandharipande (1981) equation of state.  The upper limit on the lag
is well below the maximum lag, $\bar{\omega}\sim 10$ rad s$^{-1}$
(e.g., Van Riper et al.\ 1995), sustainable by superfluid vortices
before unpinning from the crust lattice.

The rate of Joule heating caused by dissipation of the magnetic field
in the NS crust depends on the strength $B_0$ of the magnetic field
generated in the crust during or shortly after NS formation, the
density $\rho_0$ of at the crust bottom, the crust impurity parameter
$\xi$ that determines the crust conductivity, and equation of state of
the matter in the NS interior (Miralles et al.\ 1998).  Comparing our
limit with the results of Miralles et al., we can conclude that, for
instance, if $B_0\ga 10^{13}$ G, the values of $\rho_0>10^{14}$ g
cm$^{-3}$ are excluded for a plausible range of impurity parameter,
$0.001<\xi<0.01$, for both Friedman \& Pandharipande (1981) and
Pandharipande \& Smith (1975) equations of state.

The upper limit on optical flux 
gives also a constraint on the temperature and size
of hot polar cap(s) predicted by radio pulsar models (e.g., Beskin, Gurevich,
\& Istomin 1993): $T_{\rm cap} < 5 \times 10^6 R_{\rm cap,1}^{-2} d_{130}^2$~K, where
$R_{\rm cap,1}$ is the effective cap radius in km. However, this constraint is not
truly restrictive because the limiting temperature is improbably high,
$\sim 2\times 10^8$~K, for the standard
estimate of polar cap radius, $R_{\rm cap} = R (2\pi R/cP)^{1/2} \sim 0.16$~km.
X-ray observations would be more efficient for observing such small, hot polar
caps.

The limiting $B$ magnitude allows one to estimate a limit on the X-ray
flux assuming a non-thermal energy spectrum with slope $\alpha=-0.5$ as
observed for several middle-aged pulsars (e.g., Koptsevich et al.\
2001).  We obtain $F_{\rm x} < 4.6\times 10^{-15}$ erg cm$^{-2}$
s$^{-1}$ in the 0.1--2.4 keV ({\sl ROSAT}) energy band (or $<1.1\times
10^{-14}$ erg cm$^{-2}$ s$^{-1}$ in the 0.1--10 keV band). Such a flux
corresponds to an upper limit on X-ray luminosity $L_{\rm x}<
9.4\times 10^{27}$ erg s$^{-1}$ (in 0.1--2.4 keV), 
close to the value
$6\times 10^{27}$ erg s$^{-1}$ expected from the empirical
dependence $L_{\rm x}\approx 10^{-3}\dot{E}$ found for a sample of
radio pulsars observed with {\sl ROSAT} (Becker \& Tr\"umper 1997).
Even if the nonthermal radiation is a factor of a few lower than this
upper limit, it can be detected with the {\sl XMM-Newton} and {\sl
Chandra} X-ray observatories.

Clearly, it would be desirable to have a better estimate of the pulsar
distance.  The pulsar's  ecliptic latitude  is about  $10\degr$,  so a
parallax  measurement using  the NRAO  Very Long  Baseline  Array, for
example,   is  quite   feasible.  It   is  also   possible   that  the
electron-density  model  could be  improved,  giving  a more  reliable
dispersion-based distance estimate. However, this will be difficult to
achieve since the pulsar is so close.

\acknowledgements  The Guide Star Catalogue-II  is  a joint project of
the Space Telescope Science Institute and the Osservatorio Astronomico
di  Torino.  Space Telescope  Science   Institute is  operated  by the
Association   of  Universities for   Research   in Astronomy,  for the
National   Aeronautics   and   Space  Administration   under  contract
NAS5-26555.  The Australia Telescope is funded  by the Commonwealth of
Australia    for operation as  a   National  Facility  managed by  the
CSIRO. The participation of  the Osservatorio Astronomico di Torino is
supported   by the    Italian Council  for   Research in    Astronomy.
Additional support is provided by European Southern Observatory, Space
Telescope  European  Coordinating Facility,  the  International GEMINI
project and the  European Space  Agency  Astrophysics Division.    The
authors  acknowledge  the  data  analysis  facilities  provided by the
Starlink Project which is run by CCLRC on behalf of PPARC. The work of
GGP was partly supported by NASA grant NAG5-10865.

\begin{table}
\begin{tabular}{cccccc} \tableline \tableline
Date & Filter &   No.\ of exp. & Exposure(s) & seeing & airmass \\
\tableline 
2000 July 06   & $V$ & 6 & 600 & 0\farcs6 & 1.35 \\
2000 July 07   & $V$ & 6 & 600 & 0\farcs55 & 1.10\\
2000 July 26   & $U$ & 3 & 1\,800 & 0\farcs8 & 1.07 \\
2000 July 31   & $B$ & 6 & 900  & 0\farcs6 & 1.12\\
2000 August 03 & $U$ & 2 & 1\,800 & 0\farcs52& 1.02\\
\tableline 
\end{tabular}
\caption{Summary  of the  optical observations  of the  field  of PSR
J0108$-$1431 obtained with the FORS1 instrument at VLT/Antu. All the observations 
were  taken with  the same instrument configuration.  
For each  observation, the columns give the
observing date,  the filter, the number of  exposures, the integration
time per exposure, the average seeing conditions (measured directly on
the image), and the average airmass during the exposure sequence.  }
\end{table}

\begin{table}
\begin{tabular}{ccccccc} \tableline \tableline
Obj & $\Delta_\alpha \cos\delta$ & $\Delta_\delta$ & $\Delta r$ & $V$ & $B$ & $U$ \\
\tableline 
psr& 0 & 0  & 0 & $>28.0$ & $>28.6$ & $>26.4$ \\   
1  & +5.18  & +2.91  & 5.86 & 25.01 (0.02) & 25.34 (0.05) & 24.78 (0.1) \\
2  & +1.93  & --2.74 & 3.47 & 26.62 (0.10) & 27.15 (0.10) & ... \\
3  & +0.20  & --2.70 & 2.85 & 26.78 (0.15) & 27.74 (0.15) & 26.11 (0.15) \\
4  & --5.65 & +3.33  & 6.49 & 27.14 (0.15) & 27.68 (0.10) & ... \\
5  & +0.072 & +6.27  & 6.12 & 27.05 (0.15) & ...          & ... \\
6  & --5.53 & --0.53 & 5.59 & 27.92 (0.35) & ...          & 26.42 (0.25) \\
7  & +4.70  & +4.30  & 6.26 &  ...         & ...          & 26.07 (0.20) \\
8  & --4.59 & --2.16 & 5.15 &  ...         & 28.12 (0.20) & ... \\
9  & --1.06 & --3.44 & 3.75 &  ...         & 28.61 (0.25) & ...   \\
10 & --2.19 & +1.73  & 2.72 &  ...         & 28.61 (0.35) & ... \\
11 & -4.17  & --3.36 & 5.35 & 26.40 (0.3)  & ...          & ... \\

\tableline 
\end{tabular}
\caption{Point-like objects detected at $\ge 3 \sigma$ within a 
6\farcs5 radius around the pulsar position.  First column gives the object
ID according to  the labels in Figures 1--3,  second, third and fourth
columns give  the R.A.,  Dec., and the  total offsets  (in arcseconds)
from  the most  recent radio  pulsar position,  while the  three other
columns give the extinction-corrected  magnitudes
in the three filters
and the  associated errors resulting from both  statistical errors and
the uncertainties of photometric calibration.  Objects photometry
has been computed within an aperture equal to 2 times the seeing disk.
The first row gives the  $3 \sigma$ photometry upper limits within the
pulsar error ellipse. }

\end{table}


\begin{figure}
\plotone{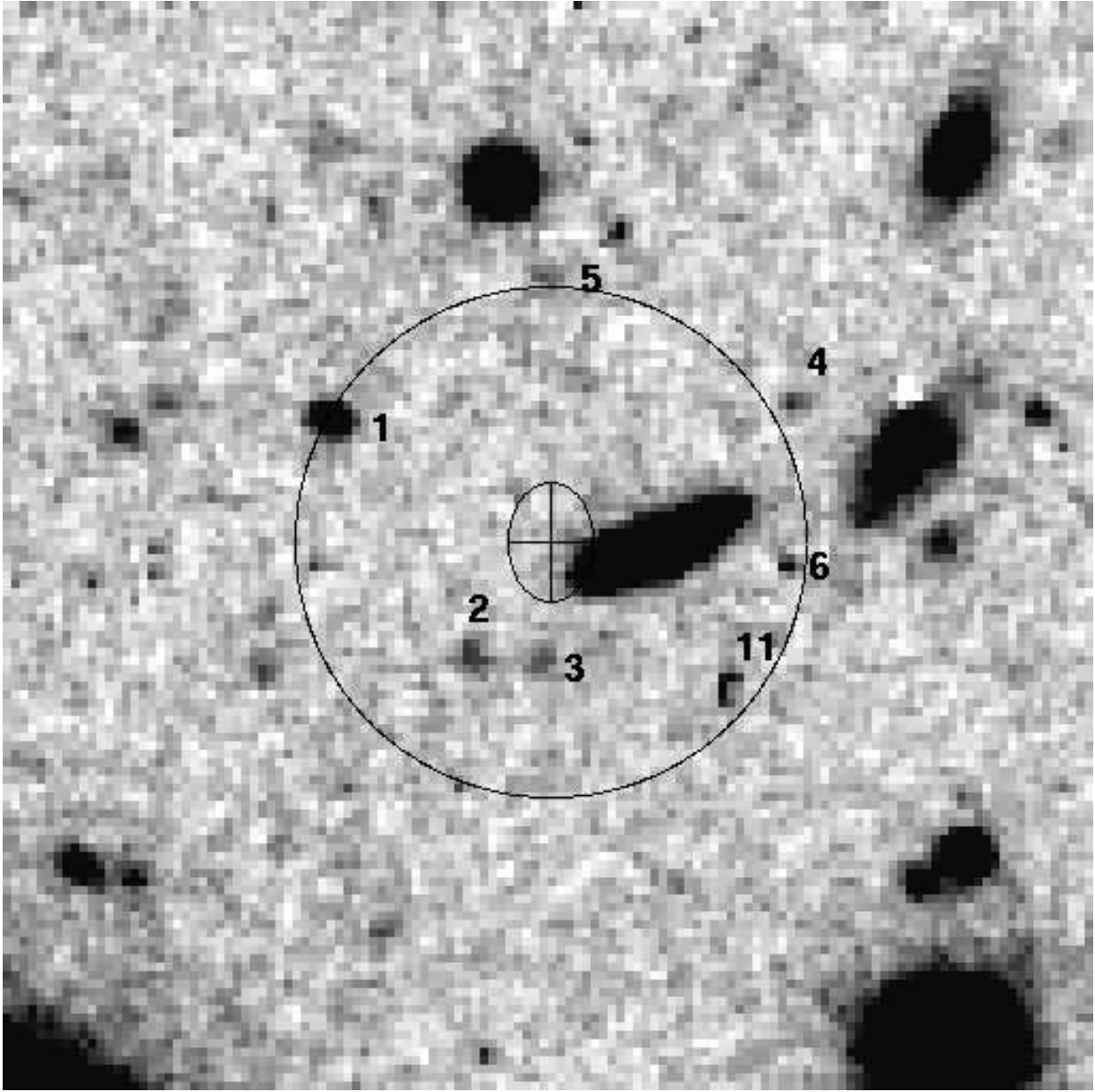}
\caption{Portion of  the $V$-band  image of the field  
around PSR J0108$-$1431
obtained with the  FORS1 instrument of the VLT-Antu  (North to the top,
East to  the left). The image, $25\farcs 6\times 25\farcs  6$, is the
average of all the available exposures (see Table  1).  The cross marks
the nominal  radio position  of the pulsar  computed according  to the
more recent radio coordinates (see  \S2.2).
The sizes of the cross arms
(minor and major semi-axes of the error ellipse), 
$3\times 0\farcs33$ in  R.A.\ and
$3\times 0\farcs46$ in Dec.,
are 3  times the overall  uncertainties of the pulsar  position 
(see  \S2.4). The radius of the circle around the pulsar position
 is $6''$. Point-like
objects  detected at  $\ge  3 \sigma$  in  at least  one passband  and
located within 
6\farcs5 radius  around the pulsar position are labelled
in Figs.\ 1--3 and listed in Table 2. \label{fig1}}
\end{figure}


\begin{figure}
\plotone{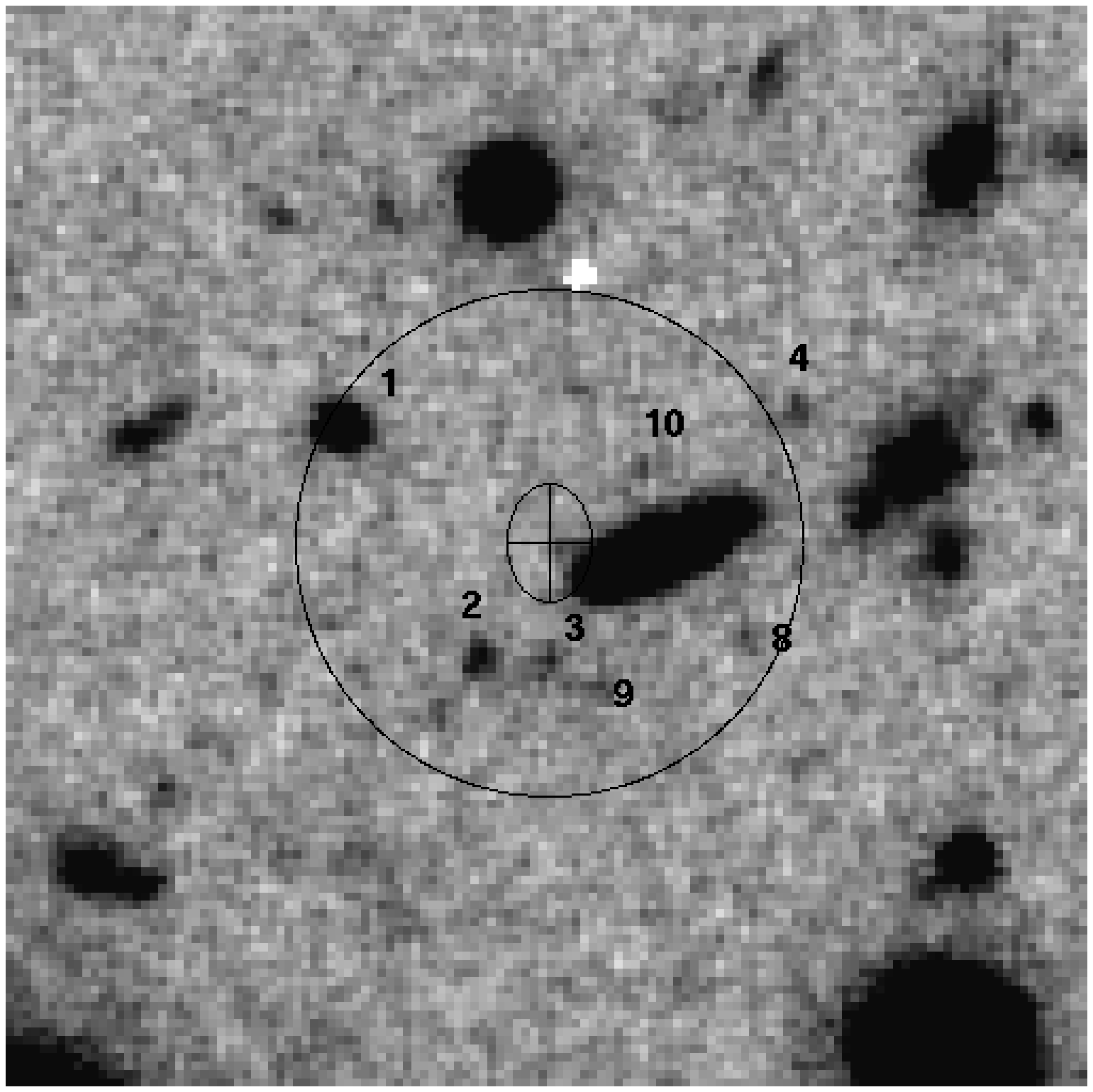}
\caption{Same as Fig.\ 1 but for the $B$ band. \label{fig2}}
\end{figure}


\begin{figure}
\plotone{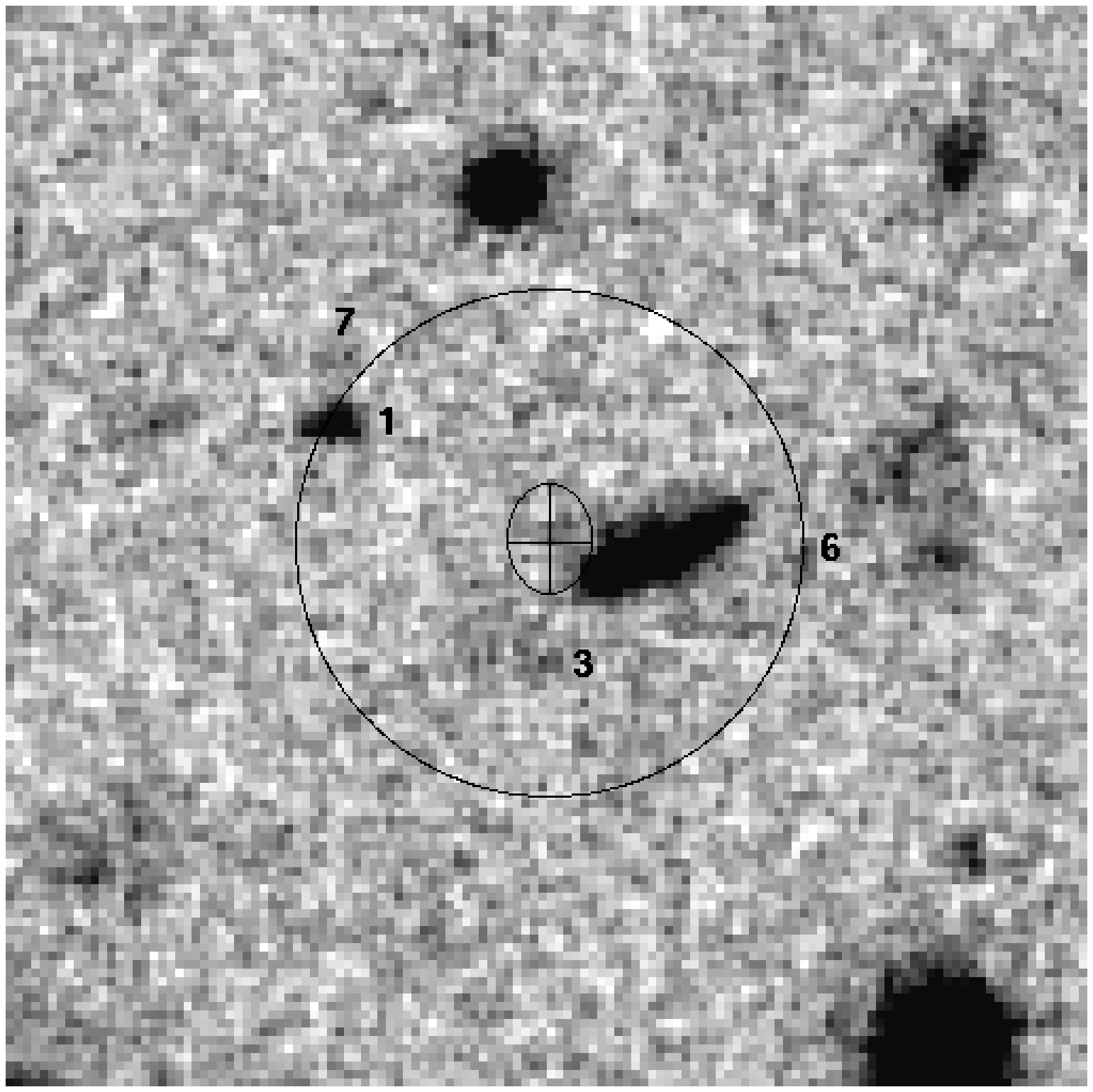}
\caption{Same as Fig.\ 1 
but for the $U$ band. \label{fig3}}
\end{figure}

\begin{figure}
\plotone{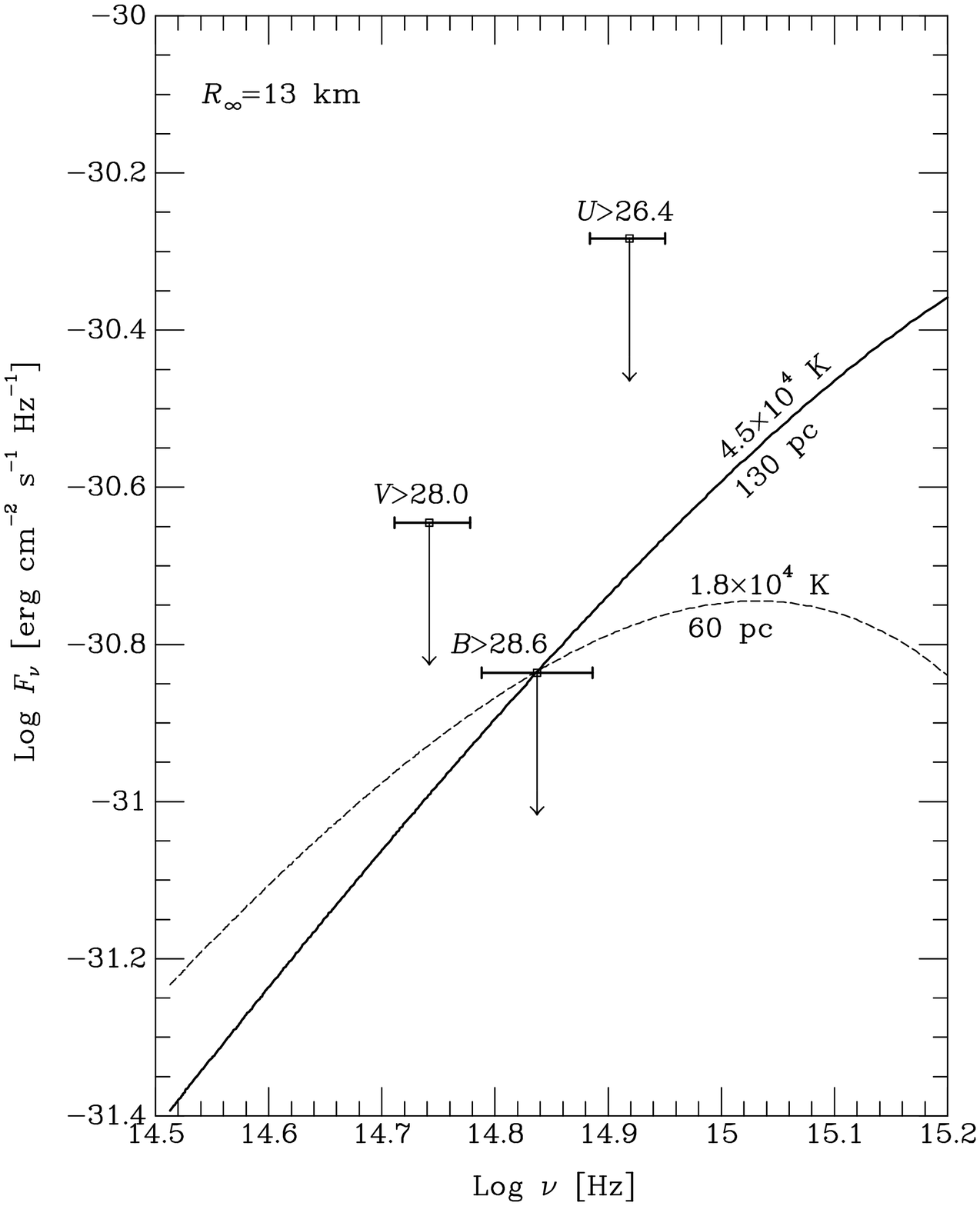}
\caption{Upper limits on the pulsar fluxes in the three optical bands
and blackbody spectra corresponding to the most stringent
$B$-band limit for the distances of 60 and 130 pc and the neutron
star radius $R_\infty = 13$ km.}
\end{figure}

\begin{thebibliography}

\bibitem{}Becker, W. \& Tr\"umper, J. 1997, A\&A, 326, 682

\bibitem{}Beskin, V.\ S., Gurevich, A.\ F., \& Istomin, Y.\ N. 1993,
Physics of Pulsar Magnetosphere (Cambridge: Cambridge Univ.\ Press)

\bibitem{}Brisken, W.\ F., Benson, J.\ M., Goss, W.\ M., \& Thorsett, S.\ E. 2002, ApJ, 571, 906

\bibitem{}D'Amico, N., Stappers, B.\ W., Bailes, M., Martin, C.\ E., Bell, J.\ F.,
Lyne, A.\ G., \& Manchester, R.\ N. 1998, MNRAS, 297, 28

\bibitem{}Dewey, R.\ J., Taylor, J.\ H., Weisberg, J.\ M., \& Stokes, G.\ H. 1985, ApJ, 294, L25

\bibitem{}Friedman, B., \& Pandharipande, V.\ R. 1981, Nucl.\ Phys.\ A. 361, 502

\bibitem{}Fukugita, M., Shimasaku, K., \& Ichikawa, T. 1995, PASP, 107, 945

\bibitem{}Koch-Miramond, L., Haas, M., Pantin, E., Podsiadlowski, P., Naylor, T.,\& Sauvage, M. 2002, A\&A, 387, 233

\bibitem{}Koptsevich, A.\ B., Pavlov, G.\ G., Zharikov, S.\ V., Sokolov, V.\ V.,
Shibanov, Yu.\ A., \& Kurt, V.\ G. 2001, A\&A, 370, 1004

\bibitem{}Korpela, E.\ J., \& Bowyer, S. 1998, AJ, 115, 2551

\bibitem{}Kurt, V.\ G., Komarova, V.\ N., Fatkhullin, T.\ A., Sokolov, V.\ V.,
Koptsevich, A.\ B., \& Shibanov, Yu.\ A. 2000,
Bull.\ Special Astrophys.\ Obs., 49, 5 (astro-ph/0005500)

\bibitem{}Larson, M.\ B., \& Link, B. 1999, ApJ, 521, 271

\bibitem{}Lorimer, D.\ R., Bailes, M., \& Harrison, P.\ A. 1997, MNRAS, 289, 592

\bibitem{}Manchester, R.\ N., et al. 1996, MNRAS, 279, 1235

\bibitem{}Mignani, R., \& Caraveo, P.\ A. 2001, A\&A, 376, 213

\bibitem{}Miralles, J.\ A., Urpin, V., \& Konenkov, D. 1998, ApJ, 503, 368

\bibitem{}Pandharipande, V.\ R., \& Smith, R.\ A. 1975, Nucl.\ Phys. A, 237, 507

\bibitem{}Pavlov, G.\ G., Stringfellow, G.\ S., \& C\'ordova, F.\ A. 1996, ApJ, 467, 370

\bibitem{}Shibazaki, N., \& Lamb, F.\ D. 1989, ApJ, 346, 808

\bibitem{}Tauris, T.\ M., et al. 1994, \apj, 428, L53

\bibitem{}Taylor, J.\ H., \& Cordes, J.\ M. 1993, ApJ, 411, 674

\bibitem{}Tsuruta, S. 1998, Phys.\ Rep. 291, 1

\bibitem{}Umeda, H., Shibazaki, N., Nomoto, K., \& Tsuruta, S. 1993, ApJ, 408, 186

\bibitem{}Van Riper, K.\ A., Link, B., \& Epstein, R.\ I. 1995, ApJ, 448, 294

\end{thebibliography}
\end{document}